\pdfoutput=1
\documentclass[11pt]{article}

\usepackage[preprint]{acl}

\usepackage{times}
\usepackage{latexsym}

\usepackage[T1]{fontenc}
\usepackage[utf8]{inputenc}

\usepackage{microtype}

\usepackage{inconsolata}

\usepackage{graphicx}

\usepackage{hyperref}
\usepackage{url}
\usepackage{soul}
\usepackage{pifont}
\usepackage{xcolor}
\usepackage{colortbl}
\usepackage[most]{tcolorbox}
\usepackage{algorithm}
\usepackage{pifont}
\usepackage[noend]{algpseudocode}
\usepackage{amsfonts}
\usepackage{fdsymbol}
\usepackage{graphicx}
\usepackage{subfig}
\usepackage{mathtools}
\usepackage{amsmath}
\usepackage{amsthm}
\usepackage{listings}
\usepackage{wrapfig}
\usepackage{makecell}
\definecolor{light_gray}{rgb}{.95,.95,.95}
\definecolor{custompurple}{RGB}{93,0,93}
\definecolor{customorange}{RGB}{255,132,6}
\definecolor{customgold}{RGB}{213,177,52}
\definecolor{customblue2}{RGB}{28,205,188}
\definecolor{no_persona_color}{RGB}{152,226,245}
\definecolor{persona_color}{RGB}{193,167,246}
\usepackage{booktabs}
\usepackage{adjustbox}
\usepackage{scalerel}
\usepackage{fontawesome5}
\newtcolorbox{policyPrompt}{
  enhanced,
  breakable,
  colback=promptbg,
  colframe=promptframe,
  boxrule=0.4pt,
  arc=1pt,
  left=6pt,
  right=6pt,
  top=6pt,
  bottom=6pt,
  fontupper=\ttfamily\small,
  coltitle=promptaccent,
  title={\textbf{Action Selection Prompt Structure}}
}
\newcommand{\llmlogo}[1]{%
    \adjustbox{valign=m}{\includegraphics[height=1.4ex]{#1}}%
}
\definecolor{myblue}{RGB}{30,92,180}
\definecolor{myRed}{RGB}{180,40,40}

\newtcolorbox{promptblock}{colback=gray!5!white,colframe=gray!75!black,
  title=Prompt Block, breakable, fontupper=\ttfamily\small}

\title{AlignUSER: Human-Aligned LLM Agents via World Models for Recommender System Evaluation}

\author{
 \textbf{Nicolas Bougie\textsuperscript{1}},
 \textbf{Gian Marconi Marconi\textsuperscript{1}},
 \textbf{Tony Yip\textsuperscript{1}},
 \textbf{Narimasa Watanabe\textsuperscript{1}}
 \\ \texttt{\{nicolas.bougie,gianmaria.marconi,tony.yip,narimasa.watanabe\}@woven.toyota}\\
\\
 \textsuperscript{1}Woven by Toyota
}

\begin{document}
\maketitle
\begin{abstract}
Evaluating recommender systems remains challenging due to the gap between offline metrics and real user behavior, as well as the scarcity of interaction data. Recent work explores large language model (LLM) agents as synthetic users, yet they typically rely on few-shot prompting, which yields a shallow understanding of the environment and limits their ability to faithfully reproduce user actions. We introduce \textsc{AlignUSER}, a framework that learns \emph{world-model-driven} agents from human interactions. Given rollout sequences of actions and states, we formalize world modeling as a next state prediction task that helps the agent internalize the environment. To align actions with human personas, we generate counterfactual trajectories around demonstrations and prompt the LLM to compare its decisions with human choices, identify suboptimal actions, and extract lessons. The learned policy is then used to drive agent interactions with the recommender system. We evaluate \textsc{AlignUSER} across multiple datasets and demonstrate closer alignment with genuine humans than prior work, both at the micro and macro levels. 
\end{abstract}

\section{Introduction}
Recommender systems (RS) are central to many online services, from e-commerce to media platforms, where they personalize content and drive user engagement \cite{li2024recent}. Despite significant progress in user preference modeling, evaluation remains a bottleneck \cite{yoon2024evaluating}. Offline metrics (e.g., nDCG, Recall) computed on static datasets dominate current evaluation practices, yet are often misaligned with online behavior once a model is deployed \cite{zhang2019deep,jannach2019measuring}. Besides, these metrics do not translate to business values such as sales or satisfaction \cite{jannach2019measuring}. On the other hand, online A/B tests offer more faithful feedback but are expensive, slow to iterate, and constrained by ethical and privacy considerations.
\begin{figure}[tbp]
    \begin{center}
        \includegraphics[width=1.0\linewidth]{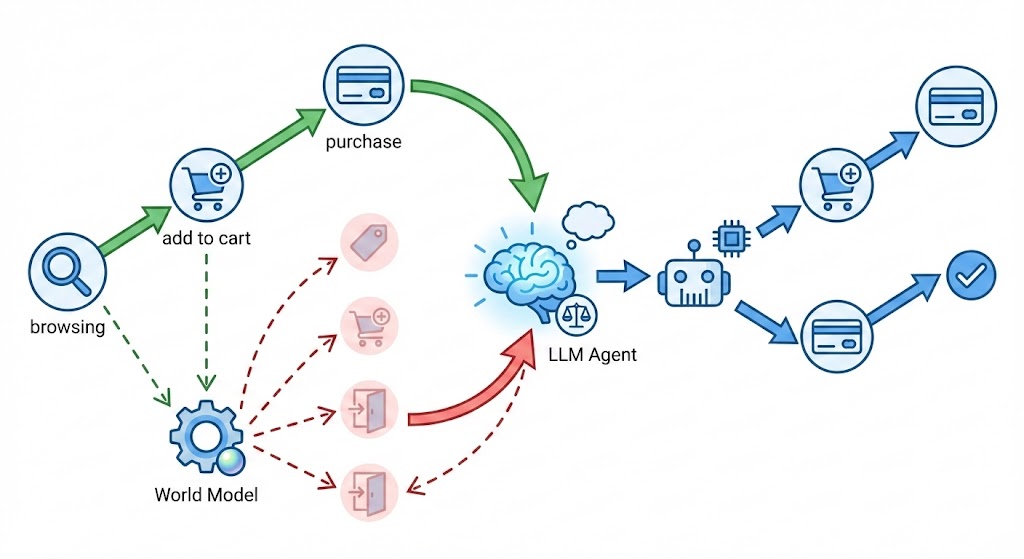}
    \caption{The \textsc{AlignUSER} framework for evaluating a recommender system by implicitly modeling a world model and exploring alternative scenarios.}
    \label{fig:overall}
\end{center}
\end{figure}
A promising alternative is to leverage LLM-based agents as synthetic users that interact with recommender systems in a simulation \cite{bougie2025simuser}. These agents can express rich preferences and feedback in natural language, potentially approximating user-level metrics such as satisfaction or perceived relevance \cite{hou2024large,zhang2023generative,huang2023recommender,wang2023recmind,yoon2024evaluating}. However, most existing approaches rely on few-shot prompting to mimic human behavior. The agent is typically asked to ``act like a typical user'' given a handful of examples \cite{wang2025opera, bougie2025simuser}, but has no explicit grasp of how the environment evolves in response to its actions. Thus, the agent gains only a superficial understanding of the world and struggles to faithfully reproduce human trajectories, especially when reasoning about long-term consequences (e.g., when to add to cart or exit). Moreover, without explicit alignment, the agent primarily projects its own intrinsic biases, rather than letting the persona realistically guide its decisions \cite{salecha2024large, 10.1145/3708319.3733685,  Bisbee_Clinton_Dorff_Kenkel_Larson_2024}.

In this paper, we postulate that agents should understand how the world works in order to faithfully replicate human actions. In light of this, given rollout trajectories (e.g., browsing, searching, adding items to the cart,...), we first pretrain the agent policy on a \emph{world-model} task that predicts the next state from a state–action pair. This task helps the agent internalize environment dynamics: what happens if it clicks on this item, goes to the next page, or decides to leave. To align agents with their human counterparts, we further generate counterfactual trajectories around demonstrations. For each state, we consider alternative actions, roll out their consequences, and prompt the LLM to compare them with the human action, identify suboptimal decisions, and extract insights to guide future choices. This reflection process yields a policy that is explicitly trained to align with human decisions while being aware of environment dynamics. At test time, the learned policy drives the agent's interactions with recommender systems. We evaluate \textsc{AlignUSER} on several datasets and show that it achieves closer alignment with humans than prior LLM-based user agents, both at the micro level, while providing more reliable guidance for RS selection than traditional offline metrics.

\section{Related Work}
\paragraph{Evaluation of recommender systems.}
Traditional recommender system evaluation predominantly relies on offline metrics such as nDCG, Recall, or RMSE computed on historical logs \cite{zhang2019deep,jannach2019measuring}. Although useful for model selection, these metrics do not directly capture user experience or business values, and their correlation with online A/B tests is often weak \cite{jannach2019measuring, bougie2025simuser}. Recent work thus explores interactive and counterfactual evaluation, including bandit simulators, user models, and causal inference techniques \cite{li2024recent}.

\paragraph{LLM-based Agents.}

Recently, LLMs have opened new possibilities for simulating human-like agents in virtual worlds~\cite{park2023generative,li2023camel,wei2022chain}. LLM-powered agents can reason, plan, and interact through natural language~\cite{wei2022chain,bougie2025generative,park2023generative, bougie2025citysim}. Several studies harness LLMs as user simulators or conversational agents in recommendation settings. RecMind \cite{wang2023recmind} and InteRecAgent \cite{huang2023recommender} propose planning and reflection mechanisms over tool-augmented agents. Agent4Rec \cite{zhang2023generative} and related work \cite{hou2024large,yoon2024evaluating} investigate generative user agents that interact with recommender models and provide ratings or textual feedback. Recently, \cite{bougie2025simuser} consider image-driven sensing and advanced reasoning modules to align agents with their human counterparts. Although these systems exhibit positive correlations with online a/b tests \cite{bougie2025simuser}, they typically treat the agent policy as a black box mapping from a textual state to an action in a single step, without an explicit model of how actions shape future states. Moreover, the agent is usually instructed to act given its persona, examples, and demographic attributes, which produces behavior reflecting the model's priors rather than genuine user patterns. 

\paragraph{World models and self-reflection.}
World models have a long history in reinforcement learning as predictors of future states and rewards \cite{ha2018world}. Recent work extends these ideas to language agents by treating states and actions as text and learning next-state predictors \cite{team2024language}. Self-reflection strategies such as STaR \cite{zelikman2022star} and more recent approaches \cite{wang2024reasoning} leverage chain-of-thought explanations to improve reasoning and robustness. Most closely related to our work, recent ``early experience'' method \cite{zhang2025agent} trains LLM agents by generating alternative trajectories and comparing expert actions to alternatives using environment feedback. We present a similar philosophy but target the alignment of user agents with human RS behavior, introduce persona-driven reflection to align persona with actions, and couple world-model-guided counterfactuals with explicit supervision from human trajectories.

\section{Problem Formulation}
We model the environment as a Markov decision process $\mathcal{M} = (\mathcal{S}, \mathcal{A}, T)$, where states $s \in \mathcal{S}$ are textual representations of pages (e.g., search results, product details, cart), and $a \in \mathcal{A}$ are actions such as \texttt{[SEARCH]}, \texttt{[CLICK]}, \texttt{[ADD\_TO\_CART]}, \texttt{[PURCHASE]}, \texttt{[RATE]}, or \texttt{[EXIT]}. We assume a dataset of $n$ human trajectories:
\begin{equation}
    \mathcal{D}_{\text{human}} = \{ (s_t^{(n)}, a_t^{(n)}, \hat{s}_{t+1}^{(n)}, p^{(n)}) \}_{t}^{n},
\end{equation}
collected from real user sessions, where $a_t^{(n)}$ denotes the human action at time $t$, and $\hat{s}_{t+1}^{(n)}$ the subsequent state. Each demonstrator is also associated with a persona $p$. Our goal is to learn a policy $\pi_\phi(a \mid s, p)$ parametrized by $\phi$, such that the trajectories it induces when interacting with the environment resemble human trajectories at both micro (step-wise action) and macro (session outcome) levels. We further assume a dataset $\mathcal{D}_{\text{rollout}}$ of experience collected either via random interactions or following a curiosity-driven strategy \cite{bougie2020skill}, $\{(s_t^{(n)}, a_t^{(n)}, \hat{s}_{t+1}^{(n)}), \ldots \}$.

\section{Method}
At its core, \textsc{AlignUSER} gains an understanding of the world by predicting next states from state–action pairs and aligns with human behaviors by comparing human actions with counterfactual examples. Following this pre-training step, the agent interacts with the recommender system. Figure~\ref{fig:overall} illustrates the overall architecture.

\subsection{World Modeling}
We first train our LLM-based policy $\pi_\phi$ to approximate the environment transition dynamics. In our study, states are represented entirely in natural language, allowing us to model next-state prediction as a standard next-token prediction objective. Inspired by prior studies on training LLMs as world models, we use next states from the rollout set $\mathcal{D}_{rollout}$ as direct training signals for the language agent’s policy $\pi_{\phi}$.

\noindent Given a state–action pair $(s_t, a_t)$ from $\mathcal{D}_{\text{rollout}}$, the model predicts the next state $s_{t+1}$ as a sequence of tokens: $\hat{s}_{t+1} \sim \pi_{\phi}(\cdot \mid s_t, a_t)$, and train $\pi_\phi$ to maximize the likelihood of the human next state $s_{t+1}^{\ast}$:
\begin{equation}
    \mathcal{L}_{\text{wm}}(\phi)
    = - \sum_{(s_t, a_t, \hat{s}_{t+1} \in \mathcal{D}_{\text{rollout}}}
    \log p_\phi(\hat{s}_{t+1} \mid s_t, a_t).
\end{equation}
For example, when browsing an e-commerce site, the model may learn to predict that clicking on a product leads to a detailed product page, whereas submitting an empty search query results in a ``no results'' state. These natural-language page descriptions act as next-state supervision, enabling the model to internalize how different user actions transform the shopping session without requiring any handcrafted supervision. 

\subsection{Human Alignment via Counterfactual Reasoning}
\label{sec:reflection}
\begin{figure}[tbp]
    \begin{center}
        \includegraphics[width=1.0\linewidth]{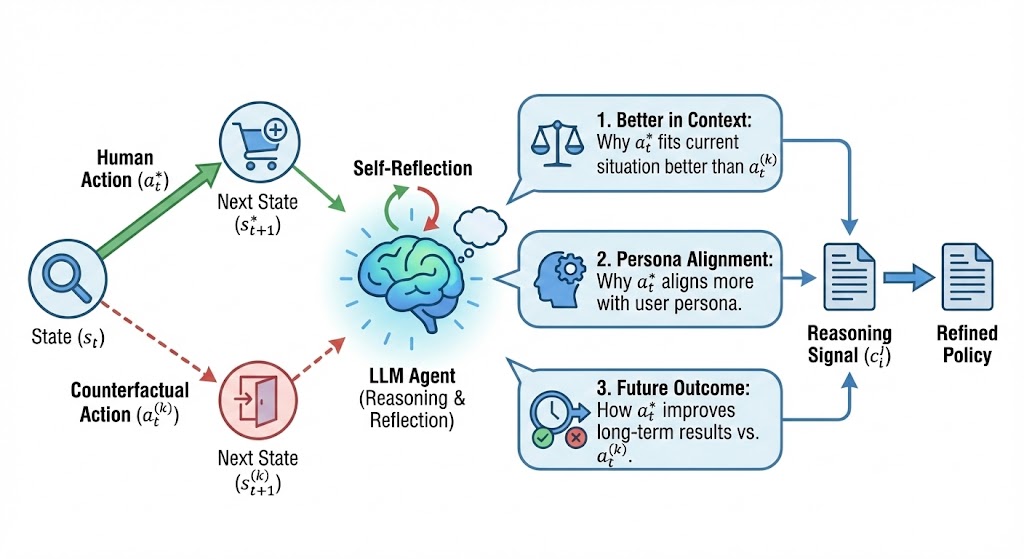}
    \caption{Counterfactual reflection from counterfactual trajectories.}
    \label{fig:reasoning}
\end{center}
\end{figure}
To align the policy with human decisions, we compare human trajectories with counterfactuals (Figure \ref{fig:reasoning}). For each human transition $(s_t, a_t, \hat{s}_{t+1},p) \in \mathcal{D}_{\text{human}}$, we sample alternative actions $\{a_t^{(1)}, \dots, a_t^{(K)}\}$ that the current policy $\pi_\phi$ considers plausible yet deviate from the demonstrated action. Given the state $s_t$, we first draw a pool of $K$ candidate actions $\{a_t^{(1)}, \dots, a_t^{(K)}\}$ from $\pi_\phi(\cdot \mid s_t, p)$ such that the generated action differs from the ground truth:
\begin{equation}
    a_t^{(k)} \sim \pi_\phi(\cdot \mid s_t, p) \quad \text{s.t.} \quad a_t^{(k)} \neq a_t
\end{equation} 
This ensures the exploration of actions that the model currently believes to be plausible, and therefore most likely to cause misalignment if left uncorrected.

\noindent We then let the agent reason on the counterfactual states, by comparing them with human state-action pairs. Given next states, we prompt the LLM to explain (1) why the human choice is better in the current context, (2) why the human choice is more aligned with its persona and preferences, (3) how the human action improves future outcomes compared to the alternatives. These explanations provide richer, transferable supervision than expert actions alone, leveraging the LLM’s strength in processing language to internalize decision principles that generalize across tasks. In practice, we prompt the model to generate a chain-of-thought $c^{j}_{t}$ explaining why the human action $a_{t}$ is preferable to the alternative $a^{j}_{t}$ based on the differences between their resulting states $\hat{s}_{t+1}$ and $s^{j}_{t}$. The prompt is designed to elicit natural language reasoning that highlights potential limitations or inefficiencies in $a^{j}_{t}$, grounded in the actual state transitions observed. This reflection is used for both environment-driven actions (e.g., click, search) and item-centric actions (e.g., like, rate). The lessons are stored in $\mathcal{D}_{\text{CR}}$. We then train the agent to jointly predict
the chain-of-thought and the expert action conditioned on the state $s_{t}$ , using a next-token prediction loss over the target sequence $(c^{j}_{t}, a_{t})$:
\begin{equation}
    \mathcal{L}_{\text{CR}} = - \sum_{(s_t, a_t, c_t^{j}, p) \in \mathcal{D}_{\text{CR}}}
    \log p_{\phi}(c_t^{j}, a_t \mid s_t, p),
\end{equation}
where $p_{\phi}$ denotes the language model’s output distribution, aligned with the agent’s policy $\pi_{\phi}$.

\noindent The overall optimization problem that is solved for learning the language agent can be expressed as:
\begin{equation}
    \mathcal{L}(\phi) = \lambda_{\text{wm}} \mathcal{L}_{\text{wm}}(\phi) + \lambda_{\text{CR}} \mathcal{L}_{\text{CR}}(\phi),
\label{eq:overall}
\end{equation}
where $\lambda_{\text{wm}}$ and $\lambda_{\text{CR}}$ are scalars that balance world-model and counterfactual terms.

\subsection{Interacting with Recommender Systems}
Once pre-trained, \textsc{AlignUSER} uses the learned policy $\pi_{\phi}$ to act as a synthetic user. Given its persona $p$, the agent interacts with the recommender system until it either purchases items or decides to terminate the session. Each agent is equipped with an episodic memory that stores its interactions with the RS. The memory is initially populated with the user's viewing and rating history. When the agent executes a new action or rates an item, the corresponding interaction is added to the episodic memory.

\noindent At each step, the policy $\pi_{\phi}$ receives a natural-language description of the current state $s_t$ (e.g., a page of recommended items), and we prompt $\pi_\phi$ to internally reason about the situation and output an action, as shown in the pseudo-prompt below:
\definecolor{promptbg}{RGB}{248,248,252}
\definecolor{promptframe}{RGB}{200,200,210}
\definecolor{promptaccent}{RGB}{40,40,60}
\begin{policyPrompt}
\small
\texttt{[STATE]}\\
\texttt{\ \ }$s_t$\\[2pt]
\texttt{[PERSONA]}\\
\texttt{\ \ }$p$\\[2pt]
\texttt{[RECENT\_HISTORY]}\\
\texttt{\ \ }$H$\\[2pt]
\texttt{[POSSIBLE\_ACTIONS]}\\
\texttt{\ \ }$a_1, a_2, \dots, a_M$\\[4pt]
\textcolor{promptaccent}{Instruction:} \texttt{Think step by step about what a careful user with this persona would do next, considering their goals, preferences, and the future consequences of each action.}\\[2pt]
\texttt{End with a single line of the form:}\\
\texttt{BEST-ACTION: <action\_token>}\\
\texttt{RATIONALE: <rationale>}
\end{policyPrompt}
The selected action is then executed in the environment (e.g., clicking an item, going to the next page, or exiting), and the process repeats until a terminal action is selected. To further enhance the ability of the agent to reason on items, we compare our vanilla \textsc{AlignUSER} with \textsc{AlignUSER+}, which integrates a graph memory, path-driven retrieval, and causal reasoning, as done in \cite{bougie2025simuser}. Namely, the agent stores its preferences in a graph-based memory and retrieves evidence to decide whether it likes or dislikes an item. Following the initial action selection $a_{\text{tent}}$, we introduce a \emph{causal reasoning} step where agents generate questions ($Q = \pi_{\phi}(a_{\text{tent}}, H, p, P_{\text{causal}})$) to validate tentative actions given recent history $H$ and prompt $P_{\text{causal}}$. For each counterfactual scenario (e.g., \textit{"What would happen if you exited now?"}), the agent estimates outcomes and adjusts its final action based on cause-effect consistency.

\section{Experiments}
\textbf{Baselines} We compare \textsc{AlignUSER} against RecAgent \cite{wang2023recagent}, Agent4Rec \cite{zhang2023generative}, and SimUSER \cite{bougie2025simuser} which represent the closest comparable methods. Some experiments involve two versions of AlignUSER: \textsc{AlignUSER} and \textsc{AlignUSER+}, in order to isolate the effects of pretraining and few-shot prompting. When possible, we also report the results of RecMind \cite{wang2023recmind}, an agent-based RS. 

\subsection{Implementation Details} 
We employ Qwen3-8B as the backbone LLM of our framework. During policy training, we generate $K=3$ counterfactual actions per state and obtain their predicted next states from the environment. The policy is trained using a weighted combination of world-model and reflection loss, following Sec.~\ref{sec:reflection}. When persona is not available, we estimate persona attributes via persona-matching, as done in SimUSER \cite{bougie2025simuser}. We investigate four real-world datasets: MovieLens-1M \cite{harper2015movielens}, Steam \cite{kang2018self}, and AmazonBook \cite{mcauley2015image}, OPeRA \cite{wang2025opera}.

\subsection{Preference Alignment}
\label{sec:beliaviability}

\begin{table*}[tb]
\centering
\begin{adjustbox}{width=\textwidth}
\begin{tabular}{cccccccccccccc}
\toprule
 & \multicolumn{4}{c}{\textbf{MovieLens}} & \multicolumn{4}{c}{\textbf{AmazonBook}} & \multicolumn{4}{c}{\textbf{Steam}} \\ 
\cmidrule(lr){2-5} \cmidrule(lr){6-9} \cmidrule(lr){10-13}
\textbf{Method(1:m)} & \textbf{Accuracy} & \textbf{Precision} & \textbf{Recall} & \textbf{F1 Score} & \textbf{Accuracy} & \textbf{Precision} & \textbf{Recall} & \textbf{F1 Score} & \textbf{Accuracy} & \textbf{Precision} & \textbf{Recall} & \textbf{F1 Score} \\ 
\midrule
RecAgent (1:1) & 0.5807 & 0.6391 & 0.6035 & 0.6205 & 0.6035 & 0.6539 & 0.6636 & 0.6587 & 0.6267 & 0.6514 & 0.6490 & 0.6499 \\
RecAgent (1:3) & 0.5077 & 0.7396 & 0.3987 & 0.5181 & 0.6144 & 0.6676 & 0.4001 & 0.5003 & 0.5873 & 0.6674 & 0.3488 & 0.4576 \\
RecAgent (1:9) & 0.4800 & 0.7491 & 0.2168 & 0.3362 & 0.6222 & 0.6641 & 0.1652 &  0.2647 & 0.5995 & 0.6732 & 0.1744 & 0.2772 \\
\midrule
Agent4Rec (1:1) & 0.6912 & 0.7460 & 0.6914 & 0.6982 & 0.7190 & 0.7276 & 0.7335 & 0.7002 & 0.6892 & 0.7059 & 0.7031 & 0.6786 \\
Agent4Rec (1:3) & 0.6675 & 0.7623 & 0.4210 & 0.5433 & 0.6707 & 0.6909 & 0.4423 & 0.5098 & 0.6505 & 0.7381 & 0.4446 & 0.5194 \\
Agent4Rec (1:9) & 0.6175 & 0.7753 & 0.2139 & 0.3232 & 0.6617 & 0.6939 & 0.2369 & 0.3183 & 0.6021 & 0.7213 & 0.1901 & 0.2822 \\
\midrule
SimUSER (1:1) & 0.7912 & 0.7976 & 0.7576 & 0.7771 & 0.8221 & 0.7969 & 0.7841 & 0.7904 & 0.7905 & 0.8033 & 0.7848 & 0.7939 \\
SimUSER (1:3) & 0.7737 & 0.8173 & 0.5223 & 0.6373 & 0.6629 & 0.7547 & 0.5657 & 0.6467 & 0.7425 & 0.8048 & 0.5376 & 0.6446 \\
SimUSER (1:9) & 0.6791 & 0.8382 & 0.3534 & 0.4972 & 0.6497 & 0.7588 & 0.3229 & 0.4530 & 0.7119 & 0.7823 & 0.2675 & 0.3987 \\
\midrule
AlignUSER (1:1) 
& 0.8203 & 0.8372 & 0.7969 & 0.8166
& 0.8432 & 0.8427 & 0.8179 & 0.8301
& 0.8138 & 0.8421 & 0.8263 & 0.8340 \\
AlignUSER (1:3) 
& 0.7994 & 0.8423 & 0.5987 & 0.6999
& 0.6843 & 0.7784 & 0.6380 & 0.7014
& 0.7641 & 0.8339 & 0.6118 & 0.7058 \\
AlignUSER (1:9) 
& 0.7061 & 0.8438 & 0.4273 & 0.5663
& 0.6648 & 0.7827 & 0.3892 & 0.5198
& 0.7284 & 0.7964 & 0.3379 & 0.4745 \\
\midrule
AlignUSER+ (1:1) 
& 0.8317 & 0.8483 & 0.8075 & 0.8274
& 0.8546 & 0.8533 & 0.8292 & 0.8416
& 0.8269 & 0.8549 & 0.8376 & 0.8462 \\
AlignUSER+ (1:3) 
& 0.8119 & 0.8532 & 0.6113 & 0.7121
& 0.6985 & 0.7915 & 0.6511 & 0.7145
& 0.7781 & 0.8461 & 0.6254 & 0.7190 \\
AlignUSER+ (1:9) 
& 0.7195 & 0.8524 & 0.4451 & 0.5848
& 0.6787 & 0.7935 & 0.4042 & 0.5356
& 0.7413 & 0.8079 & 0.3528 & 0.4922 \\
\bottomrule
\end{tabular}
\end{adjustbox}
\caption{User preference alignment across MovieLens, AmazonBook, and Steam datasets. All improvements are statistically significant ($p < 0.05$). Bold: best results for each type (1:1), (1:3) (1:9).
}
\label{table:taste_alignment}
\end{table*}

In order to appropriately respond to recommendations, synthetic users must possess a clear understanding of their own preferences. Thereby, we query the agents to classify items based on whether their human counterparts have interacted with them or not. We randomly assigned 20 items to each of 1,000 agents, with varying ratios (1:$m$ where $m \in \{1,3,9\}$) of items users had interacted with to non-interacted items. We treat this as a binary classification task. Table \ref{table:taste_alignment} shows \textsc{AlignUSER} agents accurately identified items aligned with their tastes, significantly outperforming baselines across all distractor levels (paired t-tests, 95\% confidence, $p < 0.05$). These improvements can be directly attributed to the reflection step, which allows the LLM to understand how personas relate to the agent's actions and preferences. Further gains stem from the knowledge-graph memory, as observed by comparing \textsc{AlignUSER} with \textsc{AlignUSER+}. 

\subsection{Rating Items}
\label{sec:rating}
\begin{table}[tbp]
\centering
\resizebox{1.0\linewidth}{!}{
\begin{tabular}{lcccccc}
\toprule
\textbf{Methods} & \multicolumn{2}{c}{\textbf{MovieLens}} & \multicolumn{2}{c}{\textbf{AmazonBook}} & \multicolumn{2}{c}{\textbf{Steam}} \\
 & \textbf{RMSE} & \textbf{MAE} & \textbf{RMSE} & \textbf{MAE} & \textbf{RMSE} & \textbf{MAE} \\
\midrule
MF & 1.2142 & 0.9971 & 1.2928 & 0.9879 & 1.3148 & 1.0066 \\
AFM & 1.1762 & 0.8723 & 1.3006 & 1.1018 & 1.2763 & 0.9724 \\
RecAgent  & 1.1021 & 0.7632 & 1.2587 & 1.1191 & 1.0766 & 0.9598 \\
RecMind-SI (few-shot) & 1.0651 & 0.6731 & 1.2139 & 0.9434 & 0.9291 & 0.6981 \\
Agent4Rec & 0.7612 & 0.7143 & 0.8788 & 0.6712 & 0.7577 & 0.6880 \\
SimUSER & 0.5020 & 0.4460 & 0.5676 & 0.4210 & 0.5866 & 0.5323 \\
\midrule
\rowcolor{blue!10}
\phantom{  } \textsc{AlignUSER}  & \underline{0.4693} & \underline{0.4151} & \underline{0.5130} & \underline{0.3992} & \underline{0.5344} & \underline{0.5006} \\
\rowcolor{blue!10}
\phantom{  } \textsc{AlignUSER+} & \textbf{0.4292} & \textbf{0.3871} & \textbf{0.4649} & \textbf{0.3741} & \textbf{0.4970} & \textbf{0.4829} \\
\bottomrule
\end{tabular}
}
\caption{Rating prediction performance. \textbf{Bold}: best results; \underline{underlined}: second-best. \textsc{AlignUSER}'s improvements are statistically significant ($p< 0.05$).}
\label{fig:rating_prediction}
\end{table}

A central component of recommender-system interactions is the ability to judge whether a user would like or dislike an item. We evaluate our method on this task by comparing several LLM-based agents with standard baselines, including matrix factorization (MF) \cite{koren2009matrix} and Attentional Factorization Machines (AFM) \cite{xiao2017attentional}. Results are reported in Table~\ref{fig:rating_prediction}. Across datasets, our agent consistently achieves lower error than other baselines. Other LLM approaches tend to produce larger deviations, especially on long-tail or sparsely observed items, reflecting their tendency to hallucinate plausible but incorrect ratings. In contrast, \textsc{AlignUSER} explicitly aligns the agent’s behavior with its persona during the world-model pretraining phase. This provides auxiliary signals about preference consistency and item relationships, enabling the agent to form a more coherent internal preference state before issuing a rating. 

\subsection{Human Likeliness}

\begin{table}[btp]
\centering
\resizebox{1.0\columnwidth}{!}{
\begin{tabular}{lcccc}
\toprule
 & \textbf{MovieLens} & \textbf{AmazonBook} & \textbf{Steam} & \textbf{OPeRA} \\
\midrule
RecAgent & 3.01 $\pm$ 0.14 & 3.14 $\pm$ 0.13 & 2.96 $\pm$ 0.17 & 3.05 $\pm$ 0.15 \\
Agent4Rec & 3.04 $\pm$ 0.12 & 3.21 $\pm$ 0.14 & 3.09 $\pm$ 0.16 & 3.15 $\pm$ 0.17 \\
SimUSER(persona) & 4.41$\pm$0.16 & 3.99$\pm$0.18 & 4.02$\pm$0.23 & 4.05$\pm$0.20 \\
\rowcolor{blue!10}
\phantom{  }\textsc{AlignUSER} & \underline{4.53$\pm$0.15}* & \underline{4.19$\pm$0.17}* & \underline{4.17$\pm$0.21}* & \underline{4.31$\pm$0.19}* \\
\rowcolor{blue!10}
\phantom{  }\textsc{AlignUSER+} & \textbf{4.58$\pm$0.14}* & \textbf{4.25$\pm$0.16}* & \textbf{4.22$\pm$0.20}* & \textbf{4.34$\pm$0.18}* \\
\bottomrule
\end{tabular}}
\caption{Human-likeness score evaluated by GPT-4o across recommendation domains (higher is better). *Significant improvements over best baseline ($p<0.05$).}
\label{tab:llm_evaluator}
\end{table}
To assess how closely agent trajectories resemble real user behavior, we adopt GPT-4o as an automatic evaluator, following prior evidence that LLM judges provide reliability comparable to human annotators \cite{chiang2023can}. For each interaction sequence, the evaluator assigns a score on a 5-point Likert scale, where higher values indicate stronger alignment with human-like reasoning and behavioral patterns. As shown in Table~\ref{tab:llm_evaluator}, our method achieves substantially higher human-likeness scores across all datasets. Our world modeling task reduces inconsistent behaviors and encourages the agent to evaluate how a human would behave under alternative situations. In contrast, baseline LLM agents, such as Agent4Rec, exhibit patterns that the evaluator reliably flags as non-human, including premature \texttt{[EXIT]} actions and erratic rating behavior for similar items.

\subsection{Reasoning and Persona Consistency}
\label{sec:thought_consistency}

We further measure how agents reproduce human-like reasoning and session dynamics on the OPeRA dataset~\cite{wang2025opera}, which features state-action pairs, and rationales. First, we report \textbf{thought--action consistency}, where GPT-4o compares LLM-generated and genuine rationales as \emph{coherent}, \emph{partially coherent}, or \emph{contradictory}. The consistency score is the proportion of steps labeled coherent. Second, we measure \textbf{persona--behavior consistency}, whether the actions are consistent with the stated shopping style and preferences. This targets whether the agent maintains a stable, individualized behavior pattern rather than drifting toward a generic shopper. Finally, analyze session-level metrics: number of \textbf{pages visited}, and the \textbf{purchase rate gap}, defined as the absolute difference ($\%$) between human and agent purchase frequencies.

\begin{table}[tbp]
\centering
\resizebox{1.0\linewidth}{!}{
\begin{tabular}{lcccc}
\toprule
\textbf{Method} &
\textbf{Thought--Action} &
\textbf{Persona--Behavior} &
\textbf{Pages/} &
\textbf{Purchase Rate} \\
& \textbf{Consistency (\%)} $\uparrow$ &
\textbf{Consistency (\%)} $\uparrow$ &
\textbf{Session} $\approx$ human &
\textbf{Gap (abs., \%)} $\downarrow$ \\
\midrule
Human (OPeRA) & -- & -- & 5.3 & -- \\
\midrule
Random        & 38.7 & 36.1 & 2.4 & 22.8 \\
RecAgent      & 49.5 & 46.7 & 3.5 & 16.3 \\
Agent4Rec     & 55.8 & 52.4 & 4.0 & 12.1 \\
SimUSER       & 64.3 & 61.5 & \underline{4.6} & 9.9  \\
\rowcolor{blue!10}
\textsc{AlignUSER}   & \underline{86.7} & \underline{82.4} & \textbf{5.1} & \underline{2.5} \\
\rowcolor{blue!10}
\textsc{AlignUSER+}  & \textbf{89.3} & \textbf{85.6} & \textbf{5.1} & \textbf{2.1} \\
\bottomrule
\end{tabular}}
\caption{Thought and persona consistency on OPeRA-test, together with session-level statistics. \textbf{Bold}: best result; \underline{underlined}: second best among synthetic agents.}
\label{tab:thought_consistency_opera}
\end{table}
\noindent As shown in Table~\ref{tab:thought_consistency_opera}, counterfactual reflection substantially improves internal coherence. \textsc{AlignUSER} raises thought, action consistency compared to baselines. A similar trend appears for persona, behavior consistency, indicating that the policy not only reproduces local decisions but also preserves a stable shopping style over entire sessions. Session statistics also move closer to human behavior. While RecAgent and Agent4Rec tend to under-explore the site (fewer pages than humans) and either over-purchase or under-purchase relative to human shoppers, our method produces more faithful browsing sessions.

\subsection{Action Alignment}
\begin{table*}[tbp]
\centering
\begin{tabular}{lcccc}
\toprule
\textbf{Model} &
\makecell{Action Gen.\\(Accuracy)} &
\makecell{Action Type\\(Macro F1)} & 
\makecell{Click Type\\(Weighted F1)} &
\makecell{Session Outcome\\(Weighted F1)} \\
\midrule
GPT-4.1 \llmlogo{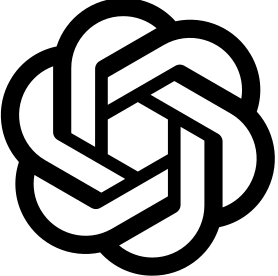} & 21.51  & 48.78 & 44.47 & 47.54 \\
\quad w/o persona & 22.06  & 45.55 & 43.45 & 58.47 \\
\qquad w/o rationale & 21.28 & 34.93 & 42.63 & 51.17 \\
\midrule
Claude-3.7 \llmlogo{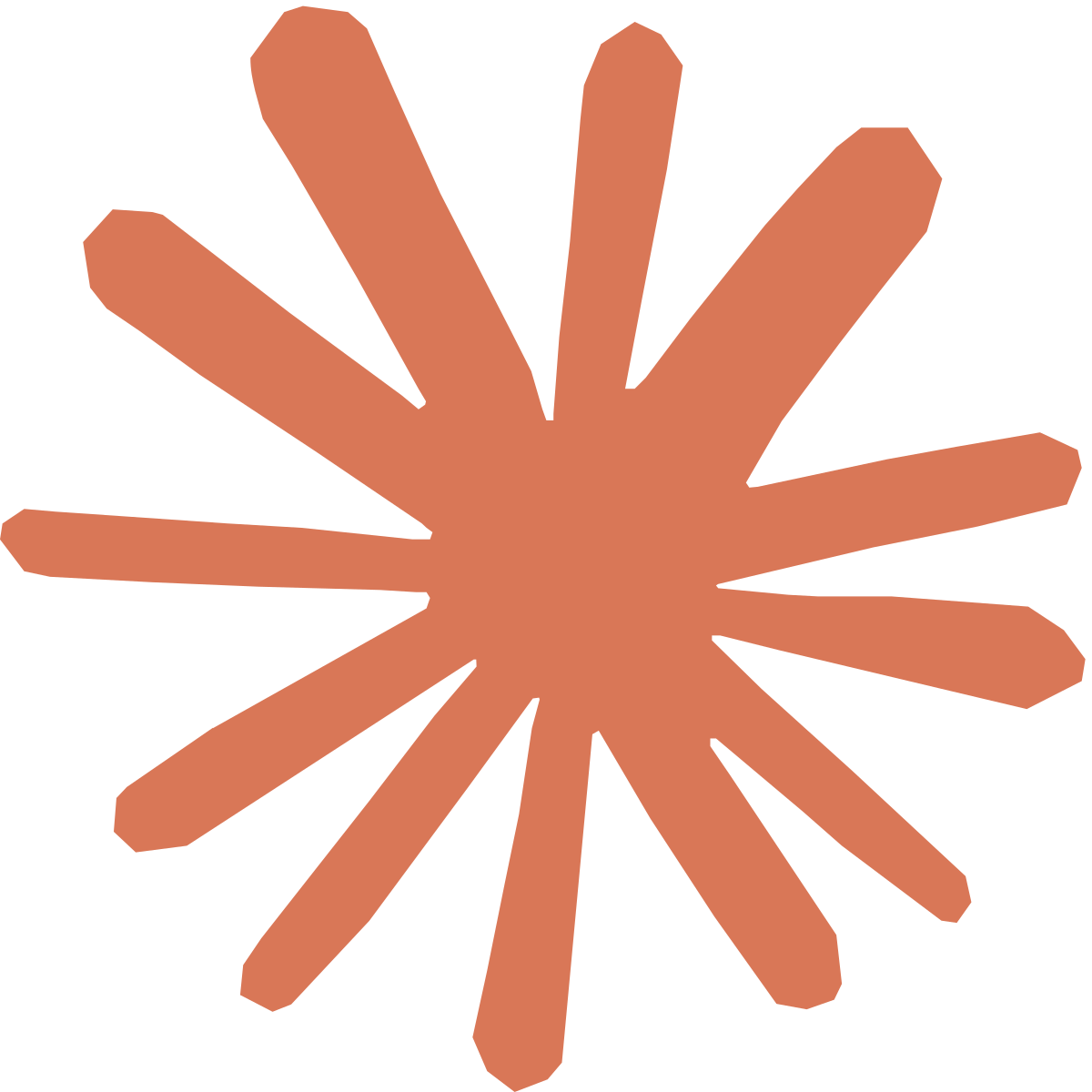} & 10.75  & 31.58 & 27.27  & 43.52 \\
\quad w/o persona & 10.75  & 25.33 & 22.76 & 43.10 \\
\qquad w/o rationale & 10.08  & 26.06 & 20.29  & 43.10 \\
\midrule
Llama-3.3 \llmlogo{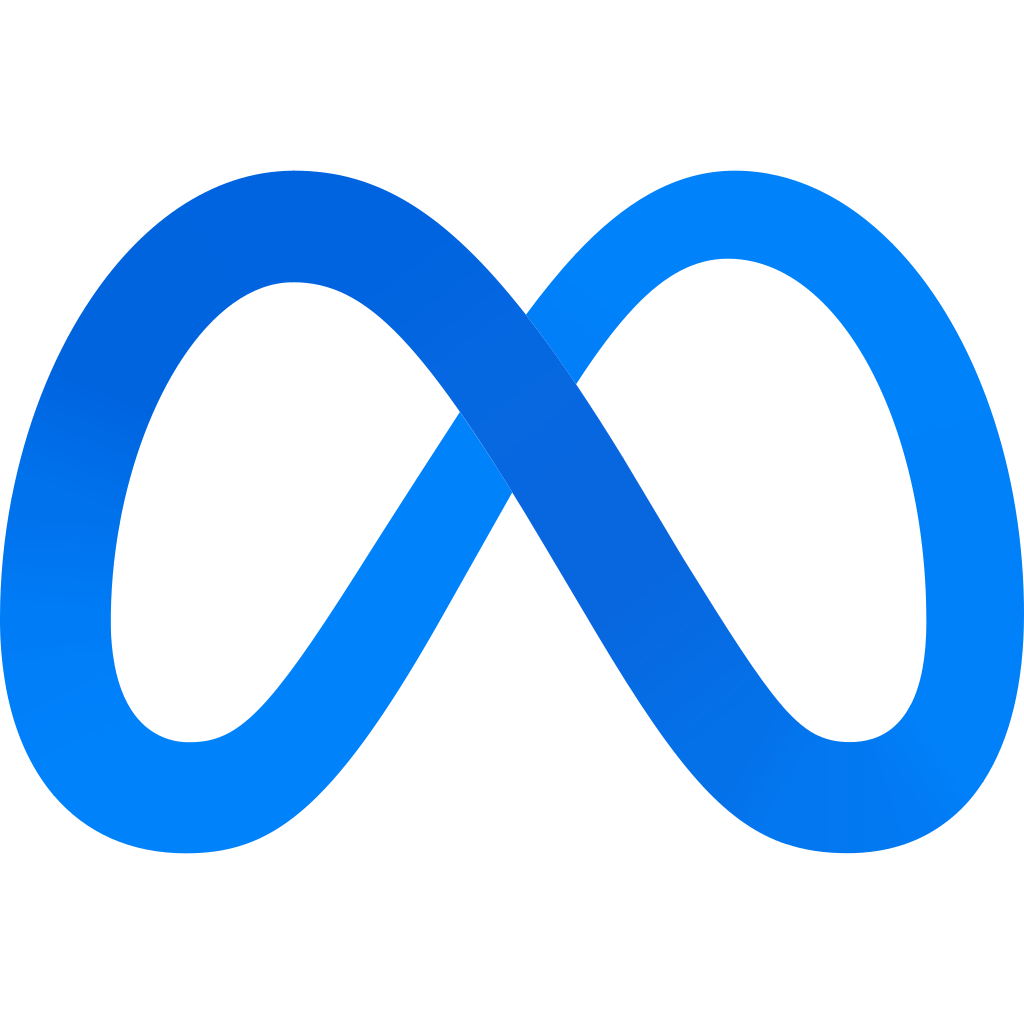} & 8.31  & 24.29 & 19.99  & 36.64 \\
\quad w/o persona & 8.31  & 23.69 & 18.59  & 33.21 \\
\qquad w/o rationale & 8.76  & 23.60 & 19.22 & 34.19 \\
\midrule
RecAgent \llmlogo{OpenAI_Logo.png} & 22.71 & 49.18 & 45.25 & 54.12 \\
Agent4Rec \llmlogo{OpenAI_Logo.png} & 23.09 & 50.05 & 46.37 & 56.70 \\
SimUSER \llmlogo{OpenAI_Logo.png} & 24.21 & 52.44 & 48.68 & 59.63 \\
\midrule
\rowcolor{blue!10}
AlignUSER \llmlogo{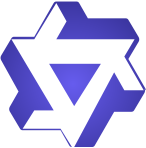}& 51.47 & 69.81 & 66.29 & 78.07 \\
\rowcolor{blue!10}
AlignUSER+ \llmlogo{Qwen_logo.png} & \textbf{52.92} & \textbf{71.94} & \textbf{66.88} & \textbf{80.52} \\
\bottomrule
\end{tabular}
\caption{
Evaluation of next-action prediction. We report four metrics: \emph{Action Generation Accuracy}, \emph{Action Type Macro F1}, \emph{Click Type Weighted F1}, and \emph{Session Outcome Weighted F1}. ``Claude-3.7'' denotes Claude-3.7-Sonnet; ``Llama-3.3'' denotes Llama-3.3-70B-Instruct. All metrics are percentages (\%).
}
\label{tab:action_predict}
\end{table*}
\begin{figure}[tbp]
    \centering
    \includegraphics[width=1.0\linewidth]{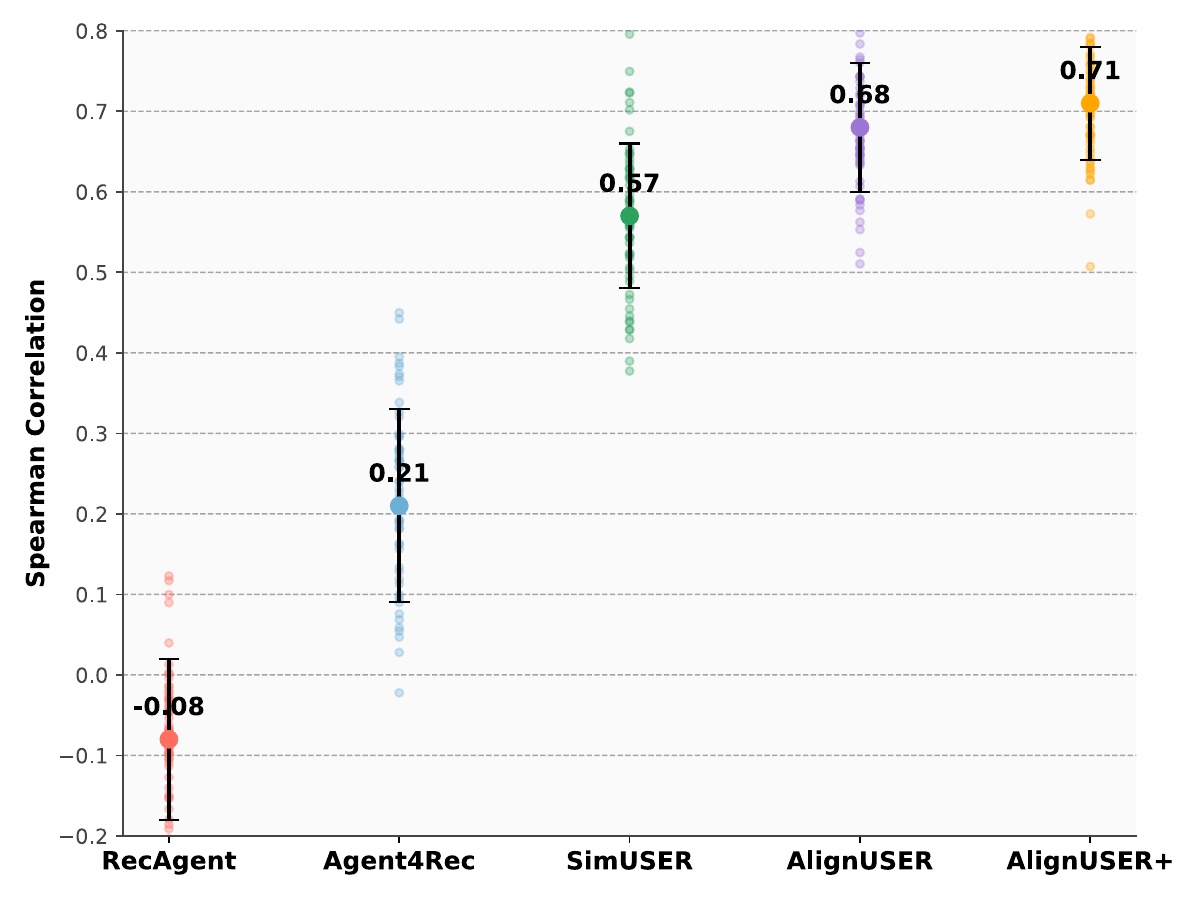} 
    \caption{Spearman correlation between estimated and actual engagement metrics. Higher values indicate better alignment with ground-truth metrics.}
    \label{fig:ab_tests}
\end{figure}

Next, we measure action alignment. We adopt an exact-match criterion: a prediction is counted as correct only if all action parameters match the ground-truth. For \texttt{click} actions, this requires matching the clicked target (e.g., the correct product or button). For \texttt{input} actions, the model must identify both the appropriate input field and the exact text entered by the user. We also assess how well each approach classifies action types. We report F1 scores for the high-level action categories \texttt{click}, \texttt{input}, and \texttt{terminate}. To assess fine-grained behavior, we further compute weighted F1 over \texttt{click} subtypes, capturing whether the model can distinguish between different click intents (e.g., \texttt{review}, \texttt{product\_link}, \texttt{purchase}). Finally, because online shopping is inherently goal-driven, we evaluate the prediction of session outcomes. We measure performance on these terminal actions using accuracy and weighted F1, which reflects how well the model captures users’ eventual decisions and long-term goals over the course of a session. As shown in Table~\ref{tab:action_predict}, \textsc{AlignUSER} surpasses prior LLM-based simulators, and \textsc{AlignUSER+} yields the strongest results, with particularly large gains in action generation accuracy and session-outcome prediction.

\subsection{Offline A/B Testing}

We further examine whether \textsc{AlignUSER} can serve as a reliable proxy for online A/B tests. We use a proprietary dataset of 55 historical A/B experiments on a large-scale food recommendation platform, each involving thousands of recommended items. Every test compares multiple recommendation strategies, with the average number of visited pages used as the primary business metric. For each strategy, we run the corresponding simulator and estimate the same engagement metric, then compute the Spearman correlation between simulated and real-world outcomes across the 55 tests. As shown in Figure~\ref{fig:ab_tests}, AlignUSER+ achieves the highest correlation with ground truth, outperforming all other baselines. Statistical tests confirm that the improvements over all baselines are significant ($p<0.05$), with AlignUSER also clearly outperforming SimUSER.

\subsection{Recommender System Evaluation}
\begin{table}[tbp]
    \centering
    \resizebox{0.9\linewidth}{!}{
    \begin{tabular}{lccccc}
    \toprule
     & $\overline{P}_{\text{view}}$ & $\overline{N}_{\text{like}}$ & $\overline{P}_{\text{like}}$ & $\overline{N}_{\text{exit}}$ & $\overline{S}_{\text{sat}}$ \\
    \midrule
    Random   & 0.295 & 3.05 & 0.247 & 2.80 & 2.60 \\
    Pop      & 0.388 & 4.15 & 0.365 & 2.95 & 3.28 \\
    MF       & 0.468 & \textbf{5.72} & 0.439 & 3.08 & 3.70 \\
    MultVAE  & \underline{0.521} & 5.31 & \textbf{0.452} & \underline{3.22} & \underline{3.82} \\
    LightGCN & \textbf{0.552} & \underline{5.49} & \underline{0.446} & \textbf{3.26} & \textbf{3.88} \\
    \bottomrule
    \end{tabular}}
    \caption{Evaluation of recommendation strategies on a recommendation task from the MovieLens dataset.}
    \label{tab:results}
\end{table}

Understanding the efficacy of various recommendation algorithms is crucial for enhancing user satisfaction. By simulating human proxies, we can better predict how users will engage with recommender systems, providing valuable interactive metrics. We compare various recommendation strategies, including most popular (Pop), matrix factorization (MF) \cite{koren2009matrix}, LightGCN \cite{he2020lightgcn}, and MultVAE \cite{liang2018variational}, using the MovieLens dataset. Upon exiting, agents rated their satisfaction on a scale from 1 to 10. Ratings above 3 were considered indicative of a \textit{like}. Metrics include average viewing ratio ($\overline{P}_{\text{view}}$), average number of likes ($\overline{N}_{\text{like}}$), average ratio of likes ($\overline{P}_{\text{like}}$), average exit page number ($\overline{N}_{\text{exit}}$), and average user satisfaction score ($\overline{S}_{\text{sat}}$). Here, rather than evaluating the proposed framework itself, we use simulated users to examine whether their interactions with recommender systems are coherent with well-established trends in the literature. Table \ref{tab:results} demonstrates that agents exhibit higher satisfaction with advanced recommendations versus random and Pop methods, consistent with real-life trends.

\section{Discussion and Limitations}
Our study demonstrates that incorporating an explicit world-modeling task and counterfactual self-reflection yields user agents that more faithfully reproduce human interaction patterns and produce more reliable evaluation signals for recommender systems. Despite these gains, several limitations remain.

\noindent The current implementation relies solely on natural-language page descriptions. Although this abstraction enables uniform modeling across domains, it omits fine-grained visual, layout, and interaction cues present in real e-commerce interfaces. Extending the framework to multimodal representations (e.g., screenshots, product images, or structured DOM states) could improve the fidelity of both the world model and the downstream policy.

\noindent Training relies on human trajectories, which provide only partial coverage of the action space. While world-model-guided counterfactual rollouts mitigate this limitation by exposing the agent to alternative transitions, the policy may still extrapolate poorly in states that are rarely visited by humans or in sequences that diverge significantly from typical browsing paths.

\noindent Our experiments focus on interaction settings with moderate temporal depth (e.g., shopping, books), where sessions typically span a few dozen steps. Deploying ser agents in long-horizon settings such as news reading, continuous mobile app use, or social media feeds may require additional components, such as persistent memory, hierarchical planning, or explicit long-term goal modeling.

\noindent Finally, the behavior of the agent inevitably inherits biases and idiosyncrasies of the underlying LLM. Although the reflection mechanism constrains deviations from human demonstrations, residual biases in preference modeling, sentiment, or perception may still surface and influence evaluation outcomes.

\section{Conclusion}
We introduced \textsc{AlignUSER}, a world-model-guided framework for learning user agents from human trajectories. By modeling environment dynamics through next-state prediction and generating counterfactuals to align actions with human decisions and personas, \textsc{AlignUSER} brings LLM-based agents closer to real user behavior. Our experiments in shopping and recommender system evaluation scenarios demonstrate improvements in action prediction, rating, and preference alignment. Our method also exhibits a positive correlation between simulated and online A/B test outcomes. We believe that synthetic users offer a promising foundation for scalable and privacy-preserving evaluation of recommender systems, and pave the way to more realistic, controllable, and interpretable agent-based simulation frameworks.

\section{Limitations}
\noindent Although \textsc{AlignUSER} achieves the highest alignment with human trajectories among the evaluated baselines, several limitations must be acknowledged. First, reproducibility is constrained by the availability of human interaction logs. A few datasets used for evaluating alignment are proprietary, limiting full transparency and replication.

\noindent Second, our approach inherits the cultural, demographic, and socioeconomic biases present in large language models. Since \textsc{AlignUSER} relies on LLM-generated reflections and counterfactual reasoning, any underlying biases in the base model may manifest as skewed interpretations of user motives or preferences. Related to this, we occasionally observe hallucinations in world-model rollouts, for example, predicting implausible next states or misinterpreting page semantics, which can propagate into the reflective policy and yield suboptimal decisions.

\noindent Third, the effectiveness of \textsc{AlignUSER} is tightly coupled to the strengths and weaknesses of the underlying LLMs. Inconsistencies in reasoning quality, brittleness under distribution shift, and occasional unfounded judgments may degrade the fidelity of simulated users, particularly in sparsely covered regions of the state-action space.

\noindent Finally, the framework integrates several interacting components, world modeling, counterfactual generation, and reflection, which can make it difficult to isolate the contribution of each module. While we provide ablation studies to partially disentangle these effects, future work is needed to better understand how different training signals and architectural choices influence alignment outcomes.

\section{Ethics Statement}
This work introduces a framework for training synthetic users to support the evaluation of recommender systems. While such agents provide clear advantages in terms of scalability, cost-effectiveness, and privacy preservation, the approach raises several ethical considerations.

\noindent Synthetic user agents trained on human logs may inadvertently reproduce or amplify demographic, cultural, or socioeconomic biases present in the underlying LLMs or in the behavioral data used for world-model learning. These biases can manifest in the agent’s simulated preferences or interaction patterns, potentially leading to misleading evaluation signals. In particular, biased reflections or hallucinated counterfactuals may privilege certain user groups or product categories, reinforcing unfairness in downstream recommender systems.

\noindent A broader concern lies around the use of realistic synthetic users as proxies for actual individuals. When such agents are used to assess or optimize RS behavior, there is a risk that system designers may over-rely on simulated outcomes, reducing the involvement of real users, domain experts, or impacted stakeholders. This is especially sensitive in domains such as e-commerce, job recommendations, or content consumption, where algorithmic decisions can influence user autonomy and exposure to information.

\noindent  Finally, the generation of counterfactual trajectories, while valuable for alignment, relies on models that may produce plausible, sounding but factually incorrect predictions. Such inaccuracies could misguide evaluation or optimization if not interpreted with caution.

\noindent We emphasize that synthetic users should complement, rather than replace, real human feedback. Responsible deployment requires transparency regarding model limitations, continuous monitoring for bias, and safeguards to prevent misuse or misinterpretation of simulation outcomes.

\bibliography{custom}

\clearpage
\appendix

\section{Experimental Setup}
\textbf{Experimental Settings.} We separate the dataset into training, validation, and test sets (80/10/10\%), using a time-based split. This ensures to reflect the temporal distribution shift that may be observed in the real world. To address privacy concerns, the name and gender are omitted. Moreover, for the sake of generality, we do not utilize user-specific information available in these datasets, relying instead on the personas identified via persona-matching \cite{bougie2025simuser}. To obtain rollouts $\mathcal{D}_{\text{rollout}}$ that cover a large space of the environment, we collect rollouts using a decaying $\epsilon$-greedy exploration policy. At episode $t$, the behavior policy selects a random action with probability $\epsilon_t$ and otherwise follows the greedy action of the current policy. We linearly anneal $\epsilon_t$ from $0.3$ to $0.05$ over $100{,}000$ episodes. $\mathcal{D}_{\text{rollout}}$ was augmented with human transitions and their counterfactual transitions. This ensures that the world model covers regions not visited by humans. Matrix factorization (MF) is utilized as the recommender model unless specified otherwise. In our simulator, agents are presented with four items $n=4$ per page and allowed to interact by viewing and rating items based on their preferences. When the output of the LLM deviated from the desired format, resulting in errors, the LLM was re-prompted with the following instruction: \texttt{You have one more chance to provide the correct answer}.

\noindent\textbf{Counterfactuals.} We generate $K=3$ counterfactual actions (excluding $a$) per state and obtain their next states from the environment. During training, we use a batch size of 16 and a learning rate of $1e^{-5}$, and train for 8 epochs. We set the loss weights to $\lambda_{\text{wm}}=1.0$ and $\lambda_{\text{CR}}=0.5$. In datasets such as MovieLens that do not include actions (e.g., \texttt{[CLICK]}, we only let the agent reflect at the item level by sampling alternative rating actions(\texttt{[1]}, \texttt{[2]},...), treating different rating values as distinct choices in the action space. In contrast, for datasets that provide interaction actions, we also generate counterfactuals at the trajectory level, enabling reflection over alternative sequences of actions and states rather than solely on isolated item-level decisions. For instance, when the supervised signal is a rating decision (e.g., MovieLens/AmazonBook), we additionally treat each discrete rating value in $\{1,2,3,4,5\}$ as a distinct action and sample alternative rating values as counterfactuals. Actions are executed in the simulation to collect $s_{t+1}$, which is then used for counterfactual reflection. 

\noindent\textbf{Preferences.} The preferences of each agent are stored in a memory, being initialized from the history of its human counterpart. When a review score for an item is greater than 4, the agent stores a memory entry in the form \texttt{I liked \{item\_name\} based on my review score of \{score\}}. For a score of 2 or below, the following format is utilized \texttt{I disliked \{item\_name\} based on my review score of \{score\}}. Neutral scores result in the entry \texttt{I felt neutral about \{item\_name\} based on my review score of \{score\}}. In all the experiments, items rated $\geq 4$ are considered as liked by the user, while items $\leq$ 2 are considered as disliked. These interactions are stored both as plain text in the episodic memory and as relationships in the knowledge graph memory. The knowledge-graph memory utilizes the same retrieval implementation and parameters as done in SimUSER \cite{bougie2025simuser}. The top-$k_{2}$ items, their attributes, and paths are returned to condition decision making (see page prompt). Namely, the titles and ratings of retrieved items are concatenated to the prompt.

\noindent\textbf{Persona.} To lay a reliable foundation for the generative agent’s subsequent interactions and evaluations, each agent has its own persona $p$. A persona $p$ encompasses a set of features that characterize the user: \textbf{age}, \textbf{personality}, and \textbf{occupation}. Personality traits are defined by the Big Five personality traits: \textit{Openness}, \textit{Conscientiousness}, \textit{Extraversion}, \textit{Agreeableness}, and \textit{Neuroticism}, each measured on a scale from 1 to 3. Along with attributes extracted from its historical data: $p \cup \{\textbf{pickiness}, \textbf{habits}\}$. \textit{pickiness} level is sampled in \{\textit{not picky}, \textit{moderately picky}, \textit{extremely picky}\} based on the user's average rating. Habits account for user tendencies in engagement, conformity, and variety \cite{zhang2023generative}. Namemly, given the average rating $\bar{R}$ of a user: $ \bar{R} = \frac{1}{N} \sum_{i=1}^{N} r_{ui}$, the pickiness level \( P(\bar{R}) \) of a user was determined based on the following thresholds:\[
P(\bar{R}) = 
\begin{cases} 
P_1 & \text{if } \bar{R} \geq 4.5 \\
P_2 & \text{if } 3.5 \leq \bar{R} < 4.5 \\
P_3 & \text{if } \bar{R} < 3.5 
\end{cases}
\]
where \( P_1 \) corresponds to \textit{not picky}, \( P_2 \) corresponds to \textit{moderately picky}, and \( P_3 \) corresponds to \textit{extremely picky}. Engagement measures the frequency and breadth of a user’s interactions with recommended items, distinguishing highly active users from those interacting with only a few items. Engagement can be mathematically expressed as: $T^{u}_{act} = \sum_{i\in \mathcal{I}} y_{ui}$, where given a user $u \in \mathcal{U}$ and an item $i \in \mathcal{I}$, the quality of the item is denoted by $R_{i} = \frac{1}{\sum_{u \in \mathcal{U}} y_{ui}} \sum_{u \in \mathcal{U}} y_{ui} \cdot r_{ui}$. $y_{ui}=0$ indicates that the user $u$ has not rated the item $i$ and inversely $y_{ui}=1$ indicates that the user has rated the item with $r_{ui} \in \{1,2,3,4,5\}$. Conformity captures how closely a user’s ratings align with average item ratings, drawing a distinction between users with unique tastes and those whose opinions closely mirror popular sentiments. For user $u$, the conformity trait is defined as: $T_{conf}^{u} = \frac{1}{\sum_{i\in \mathcal{I}} y_{ui}} \sum_{i\in \mathcal{I}} y_{ui} \cdot |r_{ui} - R_{i}|^{2}$. Variety reflects the user’s proclivity toward a diverse range of item genres or their inclination toward specific genres. The variety trait for user $u$ is formulated as: $T_{div}^{u} = |U_{i\in \{y_{ui}=1\}}g_{i}|$. 

\noindent\textbf{Interactions with Recommender Systems.} Once pre-trained, \textsc{AlignUSER} uses the learned policy $\pi_{\phi}$ to act as a synthetic user. Given a persona $p$, the agent interacts with the recommender simulator in a page-by-page manner until it selects a terminal action. Each step consists of an internal \texttt{[WATCH]}/\texttt{[SKIP]} screening over items on the current page to identify candidates consistent with the persona and memory, and selecting one environment action (e.g., navigate, click for details, or exit). The \texttt{[WATCH]}/\texttt{[SKIP]} screening is not an environment action; it is an internal decision routine to reduce the mental workload on users. During action selection, we prompt $\pi_\phi$ to internally reason about the situation and output an action. The selected action is executed in the environment (e.g., clicking an item to reveal a more detailed description, moving to the next page, or exiting), and the loop repeats until termination. In recommendation domains (MovieLens, Steam, AmazonBook), sessions terminate via \texttt{[EXIT]}; while in the web-shopping domain (OPeRA), it may include purchase-related decisions before \texttt{[TERMINATE]}. \textsc{AlignUSER+} integrates a graph memory, path-driven retrieval, and causal validation as in~ SimUSER\cite{bougie2025simuser}. The agent stores preference evidence in a graph-based memory and retrieves supporting paths to decide whether it likes or dislikes an item. Following action seelction, we introduce a causal reasoning step where agents generate questions ($Q = \pi_{\phi}(a_{\text{tent}}, H, p, P_{\text{causal}})$) to validate tentative actions. For each counterfactual scenario (e.g., \textit{"What would happen if you exited now?"}), the agent estimates outcomes and revises its final action based on cause-effect consistency.

\section{Datasets}

\noindent\textbf{MovieLens-1M.} The MovieLens-1M dataset is a widely used dataset for recommender-system research. It contains approximately 1 million movie ratings on a 1--5 star scale, provided by 6{,}040 users over 3{,}706 movies. In addition to user-item rating interactions, the dataset includes movie metadata such as titles and genre labels, as well as basic user demographic attributes, including age, gender, and occupation.

\noindent\textbf{Steam.} The Steam dataset consists of user-game interaction data collected from the Steam platform. The dataset includes user identifiers, game identifiers, and associated English-language user reviews. Game-level metadata such as titles is also provided.

\noindent\textbf{AmazonBook.} The AmazonBook dataset corresponds to a subset of the Amazon product reviews corpus restricted to the Books category. It contains user-item interactions in the form of ratings and textual reviews, along with book-level metadata such as titles and category information. 

\noindent\textbf{OPeRA.} OPeRA is a dataset designed to study and evaluate large language models for simulating human online shopping behavior. It contains real-world shopping session logs that combine user persona information collected via surveys, observations of webpage content, fine-grained user actions (e.g., clicks and navigation events), and self-reported rationales explaining users' decisions.

\section{Simulation Environment}
Our simulator mirrors real-world recommendation platforms like Netflix, or Steam, functioning in a page-by-page manner. Users are initially presented with a list of item recommendations on each page: (i) recommendations for MovieLens, Steam, and AmazonBook, and (ii) a \emph{web-shopping} pages for OPeRA. The recommendation algorithm is structured as a standalone module, allowing including any algorithm. This design features preimplemented collaborative filtering-based strategies, including
random, most popular, Matrix Factorization, LightGCN, and MultVAE. 

\noindent In \textbf{recommendation domains}, the environment displays a \emph{page} of $M$ recommended items as a single text state $s_t$. For each item, the state includes its title and an item description. The short description is either taken from available domain metadata (when present) or retrieved from the title. If the agent clicks an item, the simulator reveals a more detailed description for that item in the next state.

We format each page as:
\begin{tcolorbox}[colframe=blue!40!black, colback=white, title=Page Format (Recommendation Domains),breakable]
PAGE \{page\_number\}\\
$<$$-$ \{item\_title\} $-$$>$  $<$$-$ History ratings: \{item\_rating\} $-$$>$ $<$$-$ Summary: \{item\_description\} $-$$>$
$<$$-$ Similar items: \{similar\_items\} $-$$>$\\
$<$$-$ \{item\_title\} $-$$>$  $<$$-$ History ratings: \{item\_rating\} $-$$>$ $<$$-$ Summary: \{item\_description\} $-$$>$
$<$$-$ Similar items: \{similar\_items\} $-$$>$\\
\ldots
\end{tcolorbox}
\noindent Here, \{item\_rating\} is the agent's own historical rating when available, otherwise a dataset-derived statistic (i.e,, global mean rating). \{similar\_items\} lists retrieved neighbors from the agent's memory graph in the form \texttt{**title (rating/5)**}, and is only displayed for SimUSER and our \textsc{AlignUSER+}. The environment supports the following explicit actions: \texttt{[NEXT\_PAGE]}: advance to page $(\texttt{page\_number}+1)$. \texttt{[PREVIOUS\_PAGE]}: go back to page $(\texttt{page\_number}-1)$ when $\texttt{page\_number}>1$.\texttt{[CLICK\_ITEM:<item\_id>]}: reveal the detailed description for the selected item in the next state. \texttt{[EXIT]}: terminate the session.

\noindent In \textbf{web-shopping domains} like OPeRA, each state includes (i) page context, (ii) a product list with attributes that appear in the observation, and (iii) a list of interactive elements identified by semantic IDs.
\paragraph{State / page format (OPeRA).}
\begin{tcolorbox}[colframe=blue!40!black, colback=white, title=Page Format (OPeRA),breakable]
PAGE \{page\_number\}\\
CONTEXT: \{page\_context\}\\
PRODUCTS:\\
$<$$-$ \{product\_title\} $-$$>$ $<$$-$ Price: \{price\} $-$$>$ $<$$-$ Availability: \{availability\} $-$$>$ $<$$-$ Details: \{short\_description\} $-$$>$\\
$<$$-$ \{product\_title\} $-$$>$ $<$$-$ Price: \{price\} $-$$>$ $<$$-$ Availability: \{availability\} $-$$>$ $<$$-$ Details: \{short\_description\} $-$$>$\\
\ldots\\
INTERACTIVE ELEMENTS (semantic IDs):\\
\{semantic\_id\_1\}, \{semantic\_id\_2\}, \ldots, \{semantic\_id\_L\}
\end{tcolorbox}
\noindent Actions follow the same action space as described in OPeRA dataset \cite{wang2025opera}, extended with the navigation actions described above.

\section{Additional Experiments}
\subsection{Rating Items under Hallucination}
\begin{figure}[tbp]
    \begin{center}
        \includegraphics[width=0.9\linewidth]{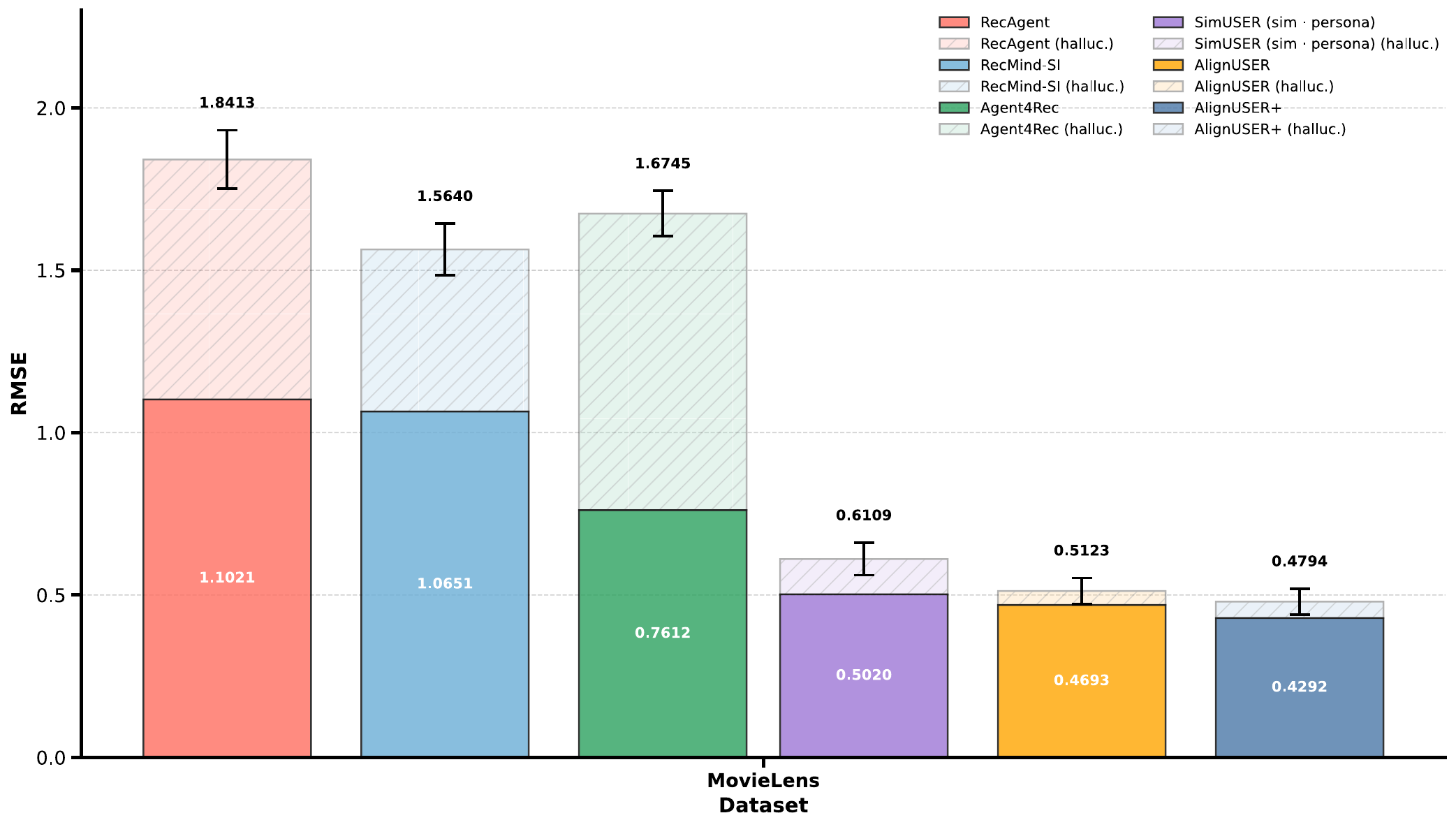} 
    \caption{Comparison of RMSE values for original (dark colors) and hallucination-affected (light colors) models for the rating task on MovieLens.}
    \label{fig:hallucination_rating}
\end{center}
\end{figure}
In this experiment, we specifically target items that are unfamiliar to the LLM, seeking to evaluate the ability of our trained agents to mitigate hallucination through their memory and alignment modules. Similarly to Section~\ref{sec:rating}, users are asked to rate movies (MovieLens), but we exclusively include items that are detected as unknown to the LLM. These items $i$ are identified by querying the LLM to classify each movie into one of 18 genres. If the LLM’s genre classification matches the actual category $g_{i}$, it indicates that the LLM is familiar with the item, and such movies are excluded from the experiment. From Figure~\ref{fig:hallucination_rating}, it is evident that while the RMSE values for all methods increase under hallucination, \textsc{AlignUSER} and \textsc{AlignUSER+} are the most robust overall. This relative robustness can be attributed to the combination of reflection and KG memory: by leveraging relationships between users, movies, and ratings from previous interactions, the agents can compare an unfamiliar movie with similar, well-known ones and anchor their predictions in familiar contexts.

\subsection{Rating Distribution}
\begin{figure}[tbp]
    \centering
    \includegraphics[width=0.90\linewidth]{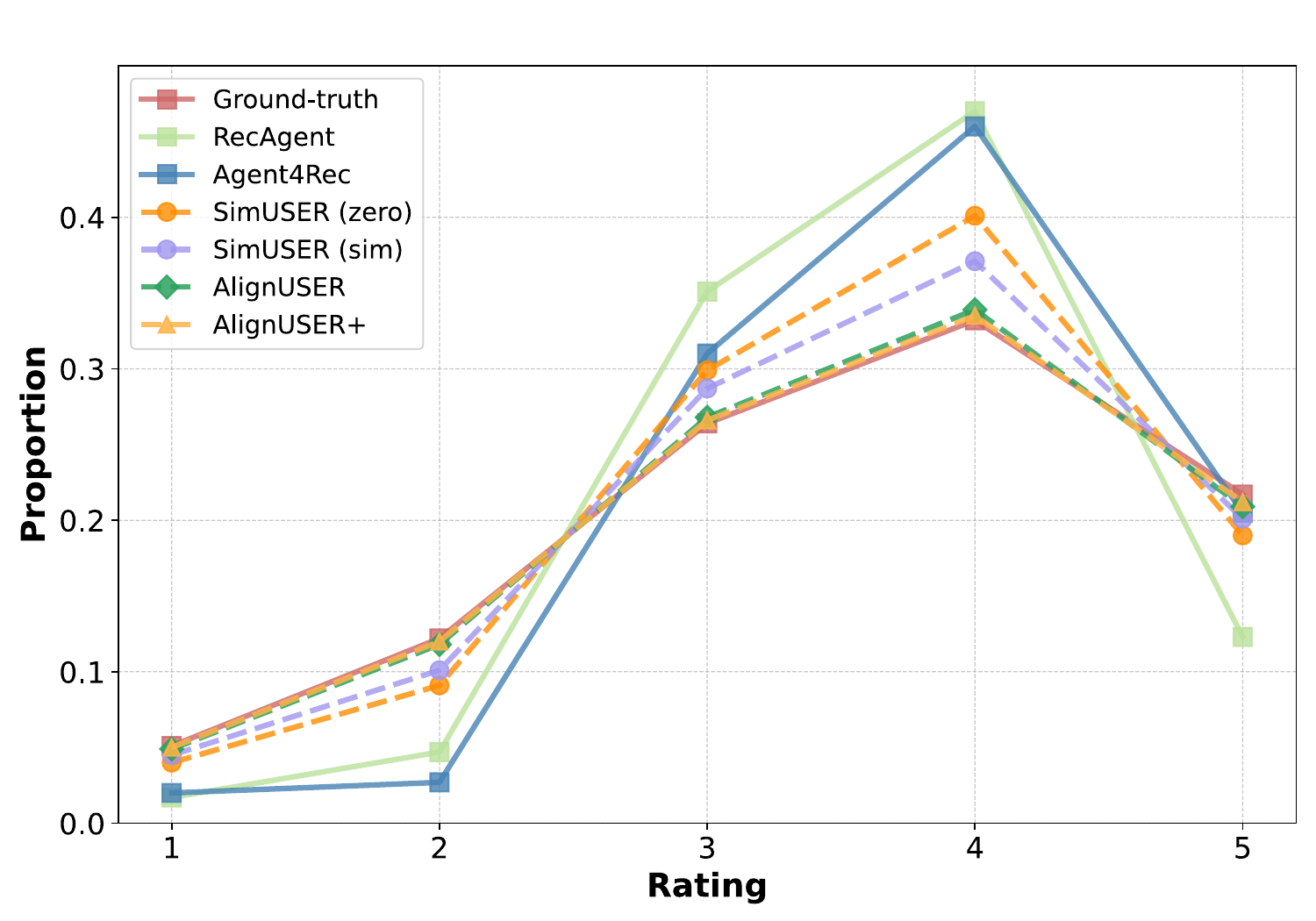}
    \caption{Comparison of rating distributions between ground-truth and human proxies.}
    \label{fig:rating_alignement}
\end{figure}

Beyond individual rating alignment, human proxies must replicate real-world behavior at the macro level. This implies ensuring that the distribution of ratings generated by the agents aligns closely with the distributions observed in the original dataset. Figure~\ref{fig:rating_alignement} presents the rating distribution from the MovieLens-1M dataset and the ratings generated by different simulators. These results reveal a high degree of alignment between the simulated and actual rating distributions, with a predominant number of ratings at 4 and a small number of low ratings (1--2). While RecAgent and Agent4Rec assign fewer low ratings than real users, SimUSER reduces this mismatch, and \textsc{AlignUSER} and \textsc{AlignUSER+} come closest to the true distribution. 

\subsection{Ablation Studies}
\begin{table}[btp]
\centering
\resizebox{1.0\columnwidth}{!}{
\begin{tabular}{lcccc}
\toprule
\textbf{Model Variant} & \textbf{MovieLens} & \textbf{AmazonBook} & \textbf{Steam} & \textbf{OPeRA} \\
\midrule
% --- Baselines (unchanged)
RecAgent & 3.01 $\pm$ 0.14 & 3.14 $\pm$ 0.13 & 2.96 $\pm$ 0.17 & 3.05 $\pm$ 0.15 \\
Agent4Rec & 3.04 $\pm$ 0.12 & 3.21 $\pm$ 0.14 & 3.09 $\pm$ 0.16 & 3.15 $\pm$ 0.17 \\
SimUSER(persona) & 4.41$\pm$0.16 & 3.99$\pm$0.18 & 4.02$\pm$0.23 & \underline{4.05$\pm$0.20} \\
\midrule
% --- Ablations
\rowcolor{gray!10}
\textsc{AlignUSER}$-$WM & 4.27$\pm$0.17 & 4.03$\pm$0.16 & 4.06$\pm$0.22 & 4.11$\pm$0.18 \\
\rowcolor{gray!10}
\textsc{AlignUSER}$-$CR & 4.12$\pm$0.18 & 3.91$\pm$0.17 & 3.87$\pm$0.21 & 3.98$\pm$0.19 \\
\rowcolor{gray!10}
\textsc{AlignUSER}$-$Persona & 4.21$\pm$0.15 & 3.95$\pm$0.16 & 3.92$\pm$0.20 & 4.02$\pm$0.18 \\
\midrule
% --- Full method
\rowcolor{blue!10}
\phantom{  }\textsc{AlignUSER} & \underline{4.53$\pm$0.15}* & \underline{4.19$\pm$0.17}* & \underline{4.17$\pm$0.21}* & \textbf{4.31$\pm$0.19}* \\
\rowcolor{blue!10}
\phantom{  }\textsc{AlignUSER+} & \textbf{4.58$\pm$0.14}* & \textbf{4.25$\pm$0.16}* & \textbf{4.22$\pm$0.20}* & \textbf{4.31$\pm$0.18}* \\
\bottomrule
\end{tabular}}
\caption{Ablation study on human-likeness task. *Significant improvements over best baseline ($p<0.05$).}
\label{tab:ablation_humanlikeness}
\end{table}

To understand the contribution of each component in our method, we perform ablations on the human-likeness evaluation across four datasets. Specifically, we remove: (1) the world-model objective ($-$WM), which prevents the agent from learning environment dynamics; (2) counterfactual reasoning ($-$CR), disabling contrastive alignment with human decisions; and (3) persona grounding ($-$Persona), which removes persona information from the policy input. Results are reported in Table \ref{tab:ablation_humanlikeness}.

\noindent Removing the world-model objective leads to a consistent drop in human-likeness, indicating that understanding environment dynamics is crucial for generating coherent interaction patterns. Eliminating counterfactual reasoning produces the sharpest decline, confirming that reflection on human–counterfactual gaps is essential for behavior alignment. Finally, ablating persona grounding reduces variability and expressiveness in simulated behavior, particularly on MovieLens and Steam, where personal preferences strongly influence item choices. The full models, \textsc{AlignUSER} and \textsc{AlignUSER+}, outperform all ablations, highlighting the importance of jointly training on world-model dynamics, persona grounding, and counterfactual reflection.

\subsection{LLM Backbone Choice}
\begin{table}[btp]
\centering
\resizebox{1.0\columnwidth}{!}{
\begin{tabular}{lcccc}
\toprule
\textbf{Backbone} & \textbf{MovieLens} & \textbf{AmazonBook} & \textbf{Steam} & \textbf{OPeRA} \\
\midrule
\textsc{AlignUSER+} (Llama-3.2-3B)   & 4.32 $\pm$ 0.18 & 4.05 $\pm$ 0.19 & 4.07 $\pm$ 0.21 & 4.01 $\pm$ 0.20 \\
\textsc{AlignUSER+} (Qwen-2.5-7B)    & 4.44 $\pm$ 0.16 & 4.13 $\pm$ 0.18 & 4.14 $\pm$ 0.20 & 4.08 $\pm$ 0.19 \\
\textsc{AlignUSER+} (Llama-3.1-8B)   & 4.52 $\pm$ 0.15 & 4.21 $\pm$ 0.17 & 4.20 $\pm$ 0.20 & 4.15 $\pm$ 0.18 \\
\rowcolor{blue!10}
\textsc{AlignUSER+} (Qwen3-8B)       & \textbf{4.58 $\pm$ 0.14} & \textbf{4.25 $\pm$ 0.16} & \textbf{4.22 $\pm$ 0.20} & \textbf{4.31 $\pm$ 0.18} \\
\bottomrule
\end{tabular}}
\caption{Human-likeness scores of \textsc{AlignUSER+} with different backbone LLMs, evaluated by GPT-4o across four recommendation domains.}
\label{tab:backbone_believability}
\end{table}
We further study the impact of the backbone LLM by swapping the base model while keeping the rest of \textsc{AlignUSER+} unchanged. As shown in Table~\ref{tab:backbone_believability}, all backbones achieve high human-likeness scores, indicating that the proposed world-model learning and counterfactual alignment are robust to the choice of underlying model. Qwen3-8B yields the strongest results overall, but the performance gaps to Llama-3.1-8B and Qwen-2.5-7B remain modest, suggesting that most of the alignment gains stem from the \textsc{AlignUSER} architecture rather than raw backbone scale.

\subsection{Running Time Analysis}
We compare the running time of AlignUSER, SimUSER, and Agent4Rec for 1,000 user interactions. While Agent4Rec and SimUSER perform API calls to GPT-4o, AlignUSER primarily performs inference with a locally served Qwen2.5 policy (vLLM). Without parallelization, Agent4Rec and SimUSER require 9.3h and 10.1h, respectively, whereas AlignUSER requires \(\approx\)0.6h using 4 GPUs. In addition, using GPT-4o pricing, 1,000 interactions costs around \$16--\$21, whereas AlignUSER costs about \$6--\$8 in GPU time (excluding one-time training).

\subsection{Next-State Prediction}
\begin{figure}[t]
    \centering
    \includegraphics[width=\linewidth]{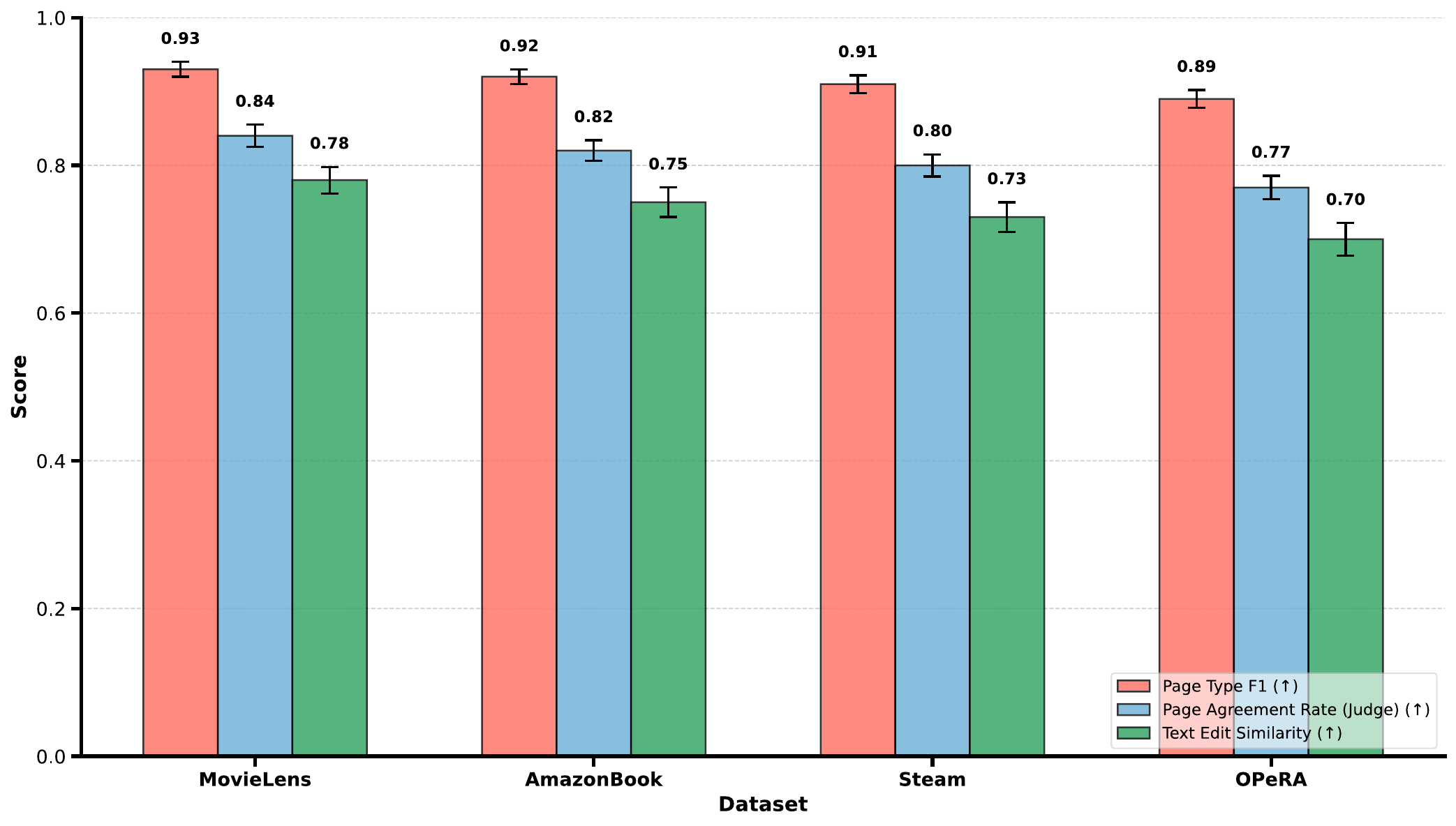}
    \caption{\textbf{Next-state prediction performance of \textsc{AlignUSER}.} Error bars indicate variation across 10 runs.}
    \label{fig:next_state_prediction}
\end{figure}
This ablation evaluates whether the next-state prediction task captures basic environment dynamics. For each held-out transition $(s_t, a_t, s_{t+1})$, we compare the predicted next state with the real one. Since next states differ in both page structure and textual content, we use three complementary metrics. (i) Page Type F1 ($\uparrow$): F1 score over coarse page categories (e.g., \texttt{browse}, \texttt{cart}), (ii) page agreement rate ($\uparrow$): a judge model assesses whether the predicted and ground-truth states describe the same page; and (iii text edit similarity. Similarity between canonicalized text representations of the two page states, based on normalized edit distance. As shown in Figure \ref{fig:next_state_prediction}, AlignUSER predicts the next page state reliably. The Judge Agreement Rate is slightly lower, which is expected because two states can be ``coherent'' even when some optional details differ (e.g., small differences in shown items). Besides, the small drop in OPeRA can be explained by the increasing state diversity and noisier content inherent in web-shopping websites.

\subsection{Sensitivity to the number of counterfactuals.}
We now study the sensitivity of \textsc{AlignUSER} to the number of counterfactual actions sampled per human step during reflection. All settings are identical to Sec.~\ref{sec:reflection} except that we vary $K \in \{0,1,2,3,5\}$, where $K{=}0$ disables counterfactual reflection (world-model pretraining only). We report next-action prediction metrics on OPeRA-test using the same protocol as Table~\ref{tab:action_predict}. Results indicate that increasing $K$ consistently improves next-action prediction, with the largest gains occurring from $K{=}0$ to $K{=}2$. Performance saturates beyond $K{=}3$, suggesting diminishing returns from additional counterfactual supervision, likely because it explores implausible actions.

\begin{table*}[tbp]
\centering
\begin{tabular}{lcccc}
\toprule
\textbf{Model} &
\makecell{Action Gen.\\(Accuracy)} &
\makecell{Action Type\\(Macro F1)} &
\makecell{Click Type\\(Weighted F1)} &
\makecell{Session Outcome\\(Weighted F1)} \\
\midrule
AlignUSER ($K{=}0$)  & 37.26 & 58.41 & 54.66 & 69.92 \\
AlignUSER ($K{=}1$)  & 44.53 & 63.72 & 60.11 & 74.35 \\
AlignUSER ($K{=}2$)  & 48.98 & 67.10 & 63.89 & 76.84 \\
\rowcolor{blue!10}
AlignUSER ($K{=}3$)  & 51.47 & 69.81 & 66.29 & 78.07 \\
AlignUSER ($K{=}5$)  & \underline{52.08} & \underline{70.12} & 66.10 & \underline{78.66} \\
\midrule
AlignUSER+ ($K{=}0$) & 39.41 & 60.07 & 56.20 & 71.58 \\
AlignUSER+ ($K{=}1$) & 46.00 & 65.31 & 61.35 & 75.32 \\
AlignUSER+ ($K{=}2$) & 50.41 & 69.02 & 64.92 & 78.11 \\
\rowcolor{blue!10}
AlignUSER+ ($K{=}3$) & 52.92 & 71.94 & 66.88 & 80.52 \\
AlignUSER+ ($K{=}5$) & \textbf{53.31} & \textbf{72.10} & \textbf{67.05} & \textbf{80.87} \\
\bottomrule
\end{tabular}
\caption{
Sensitivity to the number of counterfactual actions $K$ used for reflection training on OPeRA.
All metrics are percentages (\%). $K{=}0$ disables reflection.
}
\label{tab:k_sensitivity_action}
\end{table*}

\subsection{Persona Matching Accuracy}
\begin{table}[btp]
\centering
\begin{tabular}{lcc}
\toprule
\textbf{Metric} & \textbf{Age} & \textbf{Occupation} \\ 
\midrule
Accuracy  & 0.7184 & 0.6691 \\ 
Precision & 0.7512 & 0.6875 \\ 
Recall    & 0.7863 & 0.7386 \\ 
F1 Score  & 0.7683 & 0.7120 \\ 
\bottomrule
\end{tabular}
\caption{Performance of persona matching in predicting age and occupation utilizing MovieLens-1M.}
\label{table:persona_matching_results}
\end{table}
As personas are a central component to simulate diverse and heterogeneous users, we evaluate the effectiveness of the self-consistent persona-matching technique, being utilized in this study. Utilizing the MovieLens1 dataset, we predict the agent and occupation of users based on their interaction history. Experimental results are summarized in Table \ref{table:persona_matching_results}. Overall, persona matching turns out to be reasonably robust for enriching simulated agents with detailed backgrounds, including domains where explicit demographic data is not readily provided.

\subsection{Sensitivity to Loss Weights}
We analyze the impact of balancing world modeling and counterfactual reflection by varying the loss weights in Eq \ref{eq:overall}. When $\lambda_{\text{CR}}{=}0$, counterfactual reflection is disabled and the training objective reduces to world-model pretraining only. We report next-action prediction metrics on OPeRA-test using the same protocol as Table~\ref{tab:action_predict}. As shown in Table~\ref{tab:lambda_sensitivity_action}, introducing counterfactual reflection ($\lambda_{\text{CR}}{>}0$) yields substantial gains over world-model-only training. Performance is relatively stable around the default setting $(1.0,0.5)$.

\begin{table*}[tbp]
\centering
\begin{tabular}{lcccc}
\toprule
\textbf{Model} &
\makecell{Action Gen.\\(Accuracy)} &
\makecell{Action Type\\(Macro F1)} &
\makecell{Click Type\\(Weighted F1)} &
\makecell{Session Outcome\\(Weighted F1)} \\
\midrule
AlignUSER ($\lambda_{\text{wm}}{=}1.0,\ \lambda_{\text{CR}}{=}0$)    & 37.26 & 58.41 & 54.66 & 69.92 \\
AlignUSER ($\lambda_{\text{wm}}{=}1.0,\ \lambda_{\text{CR}}{=}0.25$) & 49.92 & 67.92 & 64.98 & 77.41 \\
\rowcolor{blue!10}
AlignUSER ($\lambda_{\text{wm}}{=}1.0,\ \lambda_{\text{CR}}{=}0.5$)  & 51.47 & 69.81 & 66.29 & 78.07 \\
AlignUSER ($\lambda_{\text{wm}}{=}1.0,\ \lambda_{\text{CR}}{=}1.0$)  & \underline{51.88} & \underline{70.03} & 66.14 & \underline{78.29} \\
AlignUSER ($\lambda_{\text{wm}}{=}2.0,\ \lambda_{\text{CR}}{=}0.5$)  & 51.21 & 69.55 & \underline{66.33} & 78.01 \\
\midrule
AlignUSER+ ($\lambda_{\text{wm}}{=}1.0,\ \lambda_{\text{CR}}{=}0$)    & 39.41 & 60.07 & 56.20 & 71.58 \\
AlignUSER+ ($\lambda_{\text{wm}}{=}1.0,\ \lambda_{\text{CR}}{=}0.25$) & 51.02 & 69.64 & 65.37 & 79.04 \\
\rowcolor{blue!10}
AlignUSER+ ($\lambda_{\text{wm}}{=}1.0,\ \lambda_{\text{CR}}{=}0.5$)  & 52.92 & 71.94 & 66.88 & 80.52 \\
AlignUSER+ ($\lambda_{\text{wm}}{=}1.0,\ \lambda_{\text{CR}}{=}1.0$)  & \underline{53.12} & \underline{72.01} & 66.74 & 80.61 \\
AlignUSER+ ($\lambda_{\text{wm}}{=}2.0,\ \lambda_{\text{CR}}{=}0.5$)  & 52.81 & 71.76 & \underline{66.92} & \underline{80.50} \\
\bottomrule
\end{tabular}
\caption{Sensitivity to loss weights $(\lambda_{\text{wm}}, \lambda_{\text{CR}})$ on OPeRA.
All metrics are percentages (\%).}
\label{tab:lambda_sensitivity_action}
\end{table*}

\section{Prompts}
\subsection{Post-Interview Prompt} 
The prompt presented to each agent for post-interview is as follows:
\begin{tcolorbox}[colframe=custompurple, colback=white, title=Post-Interview Prompt, breakable]
How satisfied are you with the recommender system you recently interacted with?\\

\textbf{\#\#\# Instructions:}\\
1. Rating: Provide a rating from 1 to 10.\\
2. Explanation: Explain the reason for your rating.\\

\textbf{\#\#\# Response Format:}\\ 
- RATING: [integer between 1 and 10]\\
- REASON: [detailed explanation]\\
\end{tcolorbox}

\subsection{Believability of Synthetic User Prompt}
In Section \ref{sec:beliaviability}, the rating prompt is modified with the following instructions:
\begin{tcolorbox}[colframe=custompurple, colback=white, title=Believably of Synthetic User Prompt, breakable]
\textbf{\#\#\# Instructions}

1. Review each \textcolor{customorange}{\{item\_type\}} in the \#\# Recommended List \#\#.\\
2. For each \textcolor{customorange}{\{item\_type\}}, classify if you have already interacted with it (``Interacted'') or if you have not (``Not Interacted'').
\end{tcolorbox}

\subsection{LLM Evaluator Prompt}
The prompt below was employed to distinguish between humans and AI-generated interactions:
\begin{tcolorbox}[colframe=custompurple, colback=white, title=LLM Evaluator Prompt, breakable]
Please evaluate the following interactions of an agent with a recommender system, and determine whether it is generated by a Large Language Model (LLM) AI or a real human:\\
\textcolor{customorange}{\{interaction logs\}}\\

Please rate on a scale of 1 to 5, with 1 being most like an AI and 5 being most like a human. 
\end{tcolorbox}

\end{document}